\documentclass[11pt,a4paper]{article}
\usepackage{epic,eepic,amsmath,latexsym,fullpage,amssymb,xcolor,amsthm}
\usepackage[normalem]{ulem}
\usepackage{ifthen,graphics,epsfig}
\usepackage{times}
\usepackage[english]{babel}
\usepackage{xspace}

\usepackage{float}
\newfloat{algorithm}{thp}{lop}
\floatname{algorithm}{Algorithm}

\begin{document}
\newlength {\squarewidth}
\renewenvironment {square}
{
\setlength {\squarewidth} {\linewidth}
\addtolength {\squarewidth} {-12pt}
\renewcommand{\baselinestretch}{0.75} \footnotesize
\begin {center}
\begin {tabular} {|c|} \hline
\begin {minipage} {\squarewidth}
\medskip
}{
\end {minipage}
\\ \hline
\end{tabular}
\end{center}
}

\newcommand{\Xomit}[1]{}
\newcommand{\Red}[1]{#1}

\newtheorem{definition}{Definition}
\newtheorem{theorem}{Theorem}
\newtheorem{lemma}{Lemma}
\newtheorem{corollary}{Corollary}
\newtheorem{property}{Property}
\newcommand{\toto}{xxx}
\newenvironment{proofT}{\noindent{\bf
Proof }} {\hspace*{\fill}$\Box_{Theorem~\ref{\toto}}$\par\vspace{3mm}}
\newenvironment{proofL}{\noindent{\bf
Proof }} {\hspace*{\fill}$\Box_{Lemma~\ref{\toto}}$\par\vspace{3mm}}
\newenvironment{proofC}{\noindent{\bf
Proof }} {\hspace*{\fill}$\Box_{Corollary~\ref{\toto}}$\par\vspace{3mm}}
\newenvironment{proofP}{\noindent{\bf
Proof }} {\hspace*{\fill}$\Box_{Property~\ref{\toto}}$\par\vspace{3mm}}

\newcommand{\BR}{{\mathit{BR}}}

\newcommand{\broadcast}{{\sf{broadcast}}}

\newcommand{\brbroadcast}{{\sf{br\_broadcast}}}
\newcommand{\brdeliver}{{\sf{br\_deliver}}}
\newcommand{\brdelivery}{{\sf{br\_delivery}}}
\newcommand{\brdelivered}{{\sf{br\_delivered}}}

\newcommand{\rbroadcast}{{\sf{r\_broadcast}}}
\newcommand{\rdeliver}{{\sf{r\_deliver}}}
\newcommand{\rdelivery}{{\sf{r\_delivery}}}
\newcommand{\rdelivered}{{\sf{r\_delivered}}}

\newcommand{\crbroadcast}{{\sf{cr\_broadcast}}}
\newcommand{\crdeliver}{{\sf{cr\_deliver}}}
\newcommand{\crdelivery}{{\sf{cr\_delivery}}}
\newcommand{\crdelivered}{{\sf{cr\_delivered}}}

\newcommand{\wwait}{{\sf{wait}}}
\newcommand{\mmax}{{\sf{max}}}

\newcommand{\hist}{{\mathit{hist}}}
\newcommand{\del}{{\mathit{del}}}
\newcommand{\sn}{{\mathit{sn}}}

\newcommand{\transfer}{{\sf{transfer}}}
\newcommand{\balance}{{\sf{balance}}}
\newcommand{\return}{{\sf{return}}}
\newcommand{\op}{{\sf{op}}}
\newcommand{\trf}{{\sf{trf}}}
\newcommand{\blc}{{\sf{blc}}}
\newcommand{\acc}{{\sf{acc}}}

\newcommand{\abort}{{\tt{abort}}}
\newcommand{\commit}{{\tt{commit}}}

\newcommand{\ttrue}{{\tt{true}}}
\newcommand{\ffalse}{{\tt{false}}}

\newcommand{\init}{{\tt{init}}}
\newcommand{\Account}{{\mathit{ACCOUNT}}}
\newcommand{\MT}{{\mathit{MT}}}
\newcommand{\Credit}{{\mathit{CREDIT}}}
\newcommand{\Balance}{{\mathit{BALANCE}}}

\newcommand{\plus}{{\sf{plus}}}
\newcommand{\minus}{{\sf{minus}}}
\newcommand{\account}{{\mathit{account}}}

\newcommand{\CAMP}{{\mathit{CAMP_{n,t}}}}
\newcommand{\BAMP}{{\mathit{BAMP_{n,t}}}}

\newcommand{\past}{{\mathit{past}}}

\newcounter{linecounter}
\newcommand{\linenumbering}{\ifthenelse{\value{linecounter}<10}{(\arabic{linecounter})}{(\arabic{linecounter})}}
\renewcommand{\line}[1]{\refstepcounter{linecounter}\label{#1}\linenumbering}
\newcommand{\resetline}[1]{\setcounter{linecounter}{0}#1}
\renewcommand{\thelinecounter}{\ifnum \value{linecounter} > 9\else \fi \arabic{linecounter}}

\newenvironment{lemma-repeat}[1]{\begin{trivlist}
\item[\hspace{\labelsep}{\bf\noindent Lemma~\ref{#1} }]}%
{\end{trivlist}}

\newenvironment{theorem-repeat}[1]{\begin{trivlist}
\item[\hspace{\labelsep}{\bf\noindent Theorem~\ref{#1} }]}%
{\end{trivlist}}

\title{\bf  Money Transfer Made Simple:
            \\ a Specification, a Generic  Algorithm, and its Proof}

\author{Alex Auvolat$^{\diamond,\dag}$,
        Davide Frey$^{\dag}$, 
        Michel Raynal$^{\dag,\star}$,
        Fran\c{c}ois Ta\"{i}ani$^\dag$  \\~\\
 $^{\diamond}${\small \'Ecole Normale Sup\'erieure, Paris, France}\\
 $^{\dag}${\small Univ Rennes, Inria, CNRS, IRISA,   35000 Rennes, France} \\
 $^{\star}${\small Department of Computing, Polytechnic University, Hong Kong}
}

\date{}
\maketitle


\begin{abstract}
It has recently been shown 
that, contrarily to a common belief, 
money transfer in the presence of faulty (Byzantine) processes
does not require strong agreement such as consensus.
This article goes one step further: namely, it first proposes a non-sequential specification of the money-transfer object, and then 
presents a generic algorithm based on a simple FIFO order between each pair
of processes that implements it. 
The genericity
dimension lies in the underlying reliable broadcast abstraction which
must be suited to the appropriate failure model. Interestingly,
  whatever the failure model, the money transfer algorithm only
  requires adding a single sequence number to its messages as control
  information.
Moreover, as a side effect of the proposed algorithm, it follows that
money transfer is a weaker problem than the construction of a
safe/regular/atomic  read/write register in the asynchronous
message-passing crash-prone model.~\\ ~\\
\noindent
    {\bf Keywords}: Asynchronous message-passing system, Byzantine
    process, Distributed computing, Efficiency, Fault tolerance, FIFO
    message order, Modularity, Money transfer, Process crash, Reliable
    broadcast, Simplicity.
\end{abstract}


\section{Introduction}  
\paragraph{Short historical perspective}
\hspace{-0.3cm}
Like  field-area or interest-rate computations, money transfers have
had a long history (see e.g.,~\cite{K72,N57}).  Roughly speaking, when
looking at money transfer in today's digital era, the issue consists
in building a software object that associates an account with each user
and provides two operations, one that allows a process to transfer
money from one account to another and one that allows a
process to read the current value of an account.  

The main issue of money transfer lies in the fact that the transfer of
an amount of money $v$ by a user to another user is conditioned to the
current value of the former user's account being at least $v$.  A
violation of this condition can lead to the problem of double spending
(i.e., the use of the same money more than once), which occurs in the
presence of dishonest processes.  Another important
issue of money transfer resides in the privacy associated with money
accounts.  This means that a full solution to money transfer must
address two orthogonal issues: synchronization (to guarantee the
consistency of the money accounts) and confidentiality/security
(usually solved with cryptography techniques).  Here, like
in closely related work~\cite{GKMPS19}, we focus on synchronization.

Fully decentralized electronic money transfer was introduced
in~\cite{N08} with the {\it Bitcoin} cryptocurrency in which there is
no central authority that controls the money exchanges issued by
users. From a software point of view, Bitcoin adopts a peer-to-peer
approach, while from an application point of view it seems to have
been motivated by the 2008 subprime crisis~\cite{R17}.

To attain its goal Bitcoin introduced a specific underlying
distributed software technology called {\it blockchain}, which can be
seen as a specific distributed state-machine-replication technique,
the aim of which is to provide its users with an object known as a
concurrent {\it ledger}.  Such an object is defined by two operations,
one that appends a new item in such a way that, once added, the item
cannot be removed, and a second operation that atomically reads the
full list of items currently appended.  Hence, a ledger builds a total
order on the invocations of its operations.  When looking at the
synchronization power provided by a ledger in the presence of
failures, measured with the consensus-number lens, it has been shown
that the synchronization power of a ledger is
$+\infty$~\cite{FKGN18,R18}.  In a very interesting way, recent
work~\cite{GKMPS19} has shown that, in a context where
each account  has a single owner
who can spend the money currently in his/her account, the consensus
number of the {\it money-transfer} concurrent object is $1$.
An owner is represented by a process in the following. 

 This is an important result, as it shows that the power of blockchain
 technology is much stronger (and consequently more costly) than
 necessary to implement money transfer\footnote{As far as we know, the
   fact that consensus is not necessary to implement money transfer
   was  stated for the first time
   in~\cite{G16}.}.  To illustrate this discrepancy,
 considering a sequential specification of the money transfer object, 
 the authors of~\cite{GKMPS19} show first that, in a failure-prone
 shared-memory system, money transfer can be implemented on top of a
 snapshot object~\cite{AADGMS93} (whose consensus number is $1$, and
 consequently can be implemented on top of read/write atomic
 registers).  Then, they appropriately modify their shared-memory
 algorithm to obtain an algorithm that works in asynchronous
 failure-prone message-passing systems. To allow the processes to
 correctly validate the money transfers, the resulting algorithm
 demands them to capture the causality relation linking money
 transfers and requires each message to carry control information
 encoding the causal past of the money transfer it carries.

\vspace{-0.2cm}
\paragraph{Content of the article}
%
The present article goes even further. It first 
presents a non-sequential specification of the money transfer
object\footnote{To our knowledge, this is the first
 non-sequential specification of the money transfer object proposed so far.}, 
and then shows that, contrarily to
what is currently accepted, the implementation of a money transfer
object does not require the explicit capture of the causality relation
linking individual money transfers.
To this end, we present a surprisingly simple yet efficient and
generic money-transfer algorithm that relies on an underlying
reliable-broadcast 
abstraction. It is efficient as it only requires a very
small amount of meta-data in its messages: in addition to money-transfer 
data, the only control information carried by the messages of
our algorithm is reduced to a single sequence number.  It is generic
in the sense that it can accommodate different failure models
\emph{with no modification}.  More precisely, our algorithm inherits
the fault-tolerance properties of its underlying reliable broadcast:
it tolerates crashes if used with a crash-tolerant reliable broadcast,
and Byzantine faults if used with a Byzantine-tolerant reliable
broadcast.

Given an $n$-process system where at most $t$ processes can be faulty,
the proposed algorithm works for $t<n$ in the crash failure model, and
$t<n/3$ in the Byzantine failure model.  This has an interesting side
effect on the distributed computability side. Namely, in the crash
failure model, money transfer constitutes a weaker problem than the
construction of a safe/regular/atomic read/write register
(where ``weaker'' means
that---unlike a  read/write register---it does
not require the ``majority of non-faulty processes'' assumption).

\vspace{-0.2cm}
\paragraph{Roadmap}
The article consists of~\ref{sec:conclusion} sections.  First,
Section~\ref{sec:models} introduces the distributed failure-prone
computing models in which we are interested, and
Section~\ref{sec:money-transfer} provides a definition of money
transfer suited to these computing models.  Then,
Section~\ref{sec:algo} presents a very simple generic money-transfer
algorithm.  Its instantiations and the associated proofs are presented
in Section~\ref{sec:algo-proof-crash} for the crash failure model and
in Section~\ref{sec:algo-proof-byz} for the Byzantine failure model.
Finally, Section~\ref{sec:conclusion} concludes the
article.\footnote{Let us note that similar ideas have been
  developed concomitantly and independently in~\cite{C-etc.20},
  which presents a money transfer system and its experimental evaluation.}

\section{Distributed Computing Models}
\label{sec:models}

\subsection{Process failure model}

\paragraph{Process model}
The system comprises a set of $n$ sequential asynchronous processes,
denoted $p_1$, ..., $p_n$\footnote{Hence the system we consider is
  static (according to the distributed computing community parlance)
  or permissioned (according to the blockchain community parlance).}.
Sequential means that a process invokes one operation at a time, and
asynchronous means that each process proceeds at its own speed, which
can vary arbitrarily and always remains unknown to the other
processes.

Two process failure models are considered.  The model parameter $t$
denotes an upper bound on the number of processes that can be faulty
in the considered model.  Given an execution $r$ (run) a process that
commits failures in $r$ is said to be faulty in $r$, otherwise it is
non-faulty (or correct) in $r$.

\paragraph{Crash failure model}
In this model, processes may crash.  A crash is a 
premature definitive halt. This means that, in the crash failure
model, a process behaves correctly (i.e., executes its algorithm)
until it possibly crashes. This model is denoted $\CAMP[\emptyset]$
(\emph{Crash Asynchronous Message Passing}).  When $t$ is restricted
not to bypass a bound $f(n)$, the corresponding restricted failure
model is denoted $\CAMP[t\leq f(n)]$.

\paragraph{Byzantine failure model}
In this model, processes can commit Byzantine
failures~\cite{LSP82,PSL80}, and those that do so are said to be
Byzantine. A Byzantine failure occurs when a process does not follow
its algorithm.  Hence a Byzantine process can stop prematurely, send
erroneous messages, send different messages to distinct processes
when it is assumed to send the same message, etc. Let us also observe
that, while a Byzantine process can invoke an operation which
generates application messages\footnote{An {\it application} message
  is a message sent at the application level, while an {\it
    implementation} is low level message used to ensure the correct
  delivery of an application message.}  it can also ``simulate'' this
operation by sending fake implementation messages that give their
receivers the illusion that they have been generated by a correct
sender.  However, we assume that there is no Sybil attack like most
previous work on byzantine fault tolerance
including~\cite{GKMPS19}.\footnote{As an example, a Byzantine process
  can neither spawn new identities, nor assume the identity of
  existing processes.}

As previously, the notations $\BAMP[\emptyset]$ and $\BAMP[t\leq f(n)]$
(\emph{Byzantine Asynchronous Message Passing})
are used to refer to the corresponding Byzantine failure models.

\subsection{Underlying complete point-to-point network}
The set of  processes communicate through an underlying message-passing
point-to-point network in which there exists a bidirectional channel
between any pair of processes. Hence, when a process receives a
message, it knows which process sent this message.  For
simplicity, in writing the algorithms, we assume that a process can
send messages to itself.

Each channel is reliable and asynchronous. Reliable means that
a channel does not lose, duplicate, or corrupt messages.
Asynchronous means that the transit delay of each   message
is finite but arbitrary.  Moreover, in the case of the Byzantine
failure model, a Byzantine process can read the content of the messages
exchanged through the channels,  but cannot modify their content. 

To make our algorithm as generic and simple as possible,
Section~\ref{sec:algo} does not present it in terms of low-level
send/receive operations\footnote{Actually the send and receive
  operations can be seen as ``machine-level'' instructions provided by
  the network.}  but in terms of a high-level communication
abstraction, called {\it reliable broadcast}
(e.g.,~\cite{B87,CGR11,HT94,IR16,R18}).  The definition of this
communication abstraction appears in
Section~\ref{sec:algo-proof-crash} for the crash failure model and
Section~\ref{sec:algo-proof-byz} for the Byzantine failure model.  It
is important to note that the previously cited reliable broadcast
algorithms do not use sequence numbers.  They only use different types
of implementation messages which can be encoded with two bits.

\section{Money Transfer: a Formal Definition}
\label{sec:money-transfer}

\paragraph{Money transfer: operations}
From an abstract point of view, a money-transfer object can be seen as
an abstract array $\Account[1..n]$ where $\Account[i]$ represents the
current value of $p_i$'s account.  This object provides the processes
with two operations denoted $\balance()$ and $\transfer()$, whose
semantics are defined below.  The transfer by a process of the amount
of money $v$ to a process $p_j$ is represented by the pair $\langle
j,v\rangle$.  Without loss of generality, we assume that a process
does not transfer money to itself. It is assumed that each
$\Account[i]$ is initialized to a non-negative value denoted
$\init[i]$. It is assumed the array $\init[1..n]$ is initially known
by all the processes.\footnote{It is possible to initialize some
  accounts to negative values.  In this case, we must assume $pos > neg$,
where  $pos$ (resp., $neg$) is the sum of
all the positive (resp., negative) initial values.}

Informally, when $p_i$ invokes $\balance(j)$ it obtains a value
(as defined below) of $\Account[j]$, and when it invokes the
transfer $\langle j,v\rangle$, the amount of money $v$ is moved from
$\Account[i]$ to $\Account[j]$. If the transfer succeeds, the operation
 returns ${\commit}$, if it fails it  returns ${\abort}$.

\paragraph{Histories}
The following notations and definitions
are inspired  from~\cite{ANBHK95}. 
\begin{itemize}
\vspace{-0.2cm}
\item
A local execution history (or local history) of a process $p_i$,
denoted $L_i$, is a sequence of operations $\balance()$ and
$\transfer()$ issued by $p_i$.  If an operation $\op1$ precedes an
operation $\op2$ in $L_i$, we say that ``$\op1$ precedes $\op2$ in
process order'', which is denoted $\op1 \rightarrow_i\op2$.
\vspace{-0.3cm}
\item
An execution history (or history) $H$~is a set of $n$ local
histories, one per process, $H=(L_1,\cdots,L_n)$.
\vspace{-0.3cm}
\item
  A serialization $S$ of a history $H$ is a sequence  that contains
  all the operations of $H$ and respects the process order $\rightarrow_i$
  of each process $p_i$.
\vspace{-0.2cm}
\item
Given a history $H$ and a process $p_i$, let 
$A_{i,T}(H)$ denote the history $(L_1',...,L_n')$ such that 
\begin{itemize}
\vspace{-0.3cm}
\item  $L_i'=L_i$, and 
\vspace{-0.2cm}
\item  For any $j\neq i$: $L_j'$ contains only the 
 transfer operations of $p_j$. 
\end{itemize}
\end{itemize}

\vspace{-0.7cm}
\paragraph{Notations}
\begin{itemize}
\vspace{-0.2cm}
\item
An operation $\transfer(j,v)$ invoked by $p_i$ is denoted  $\trf_i(j,v)$.
\vspace{-0.2cm}
\item
An invocation of $\balance(j)$ that returns the value $v$ is denoted
$\blc(j)/v$.
\vspace{-0.2cm}
\item Let $H$ be a set of operations. 
\begin{itemize}
\vspace{-0.2cm}
\item $\plus(j,H)= \Sigma_{\trf_k(j,v)\in H}~~ v$
  (total of the money given to $p_j$ in $H$).  
\vspace{-0.1cm}
\item $\minus(j,H)= \Sigma_{\trf_j(k,v)\in H}~~ v$
  (total of the money given by $p_j$ in $H$). 
\vspace{-0.1cm}
\item $\acc(j,H)= \init[j] + \plus(j,H)-\minus(j,H)$
  (value of $\Account[j]$ according to $H$). 
\end{itemize}
\vspace{-0.1cm}
\item
Given a history $H$ and a process $p_i$, let
$S_i$ be a  serialization of $A_{i,T}(H)$
(hence, $S_i$ respects the $n$ process orders defined by $H$).
Let  $\rightarrow_{S_i}$ denote the total order defined by $S_i$. 

\end{itemize}

\paragraph{Money-transfer-compliant  serialization}
A serialization $S_i$ of  $A_{i,T}(H)$ is money-transfer compliant
(MT-compliant) if:
\begin{itemize}
\vspace{-0.3cm}
\item For any operation $\trf_j(k,v)\in S_i$, we have \\
\centerline{$v\leq \acc(j,\{ \op \in S_i~|~ \op\rightarrow_{S_i} \trf_j(k,v)\}$), and}
\vspace{-0.7cm}
\item For any operation $\blc(j)/v \in S_i$, we have \\
\centerline{
    $v= \acc(j, \{ \op \in S_i~|~ \op\rightarrow_{S_i} \blc(j)/v\}$).}
\end{itemize}
MT-compliance is the key concept at the basis of the definition of a
money-transfer object.  It states that it is possible to 
associate each process $p_i$ with 
a total order $S_i$ in which (a) each of its invocations of
$\balance(j)$ returns a value $v$ equal to $p_j$'s account's
current value according to $S_i$, and (b) processes transfer only
money that they have.

Let us observe that the common point among the serializations
$S_1$, ..., $S_n$ lies in the fact that each  process sees all the transfer
operations of any other  process $p_j$ in the order
they have been produced (as defined by $L_j$), and  
sees its own transfer and balance operations
in the order it produced them  (as defined by $L_i$).

\paragraph{Money transfer in $\CAMP[\emptyset]$}
Considering the $\CAMP[\emptyset]$ model, 
a money-transfer object is an object
that provides the processes with $\balance()$
and $\transfer()$ operations and is such that, for each of its
executions, represented by the corresponding history $H$, we have:
\begin{itemize}
\vspace{-0.1cm}
\item All the operations invoked by correct processes terminate.
\vspace{-0.2cm}
\item For any correct process $p_i$, there is 
  an MT-compliant serialization $S_i$ of  $A_{i,T}(H)$, and 
\vspace{-0.2cm}
\item For any faulty process $p_i$, there is a history
  $H'=(L_1',...,L_n')$ where (a) $L_j'$ is a prefix of $L_j$ 
  for any $j\neq i$,  and (b) $L_i'=L_i$, and there is
  an MT-compliant serialization of  $A_{i,T}(H')$. 
\end{itemize}
An algorithm implementing a money transfer object is correct
in $\CAMP[\emptyset]$  if it produces only executions as defined above. 
We then say that the algorithm is MT-compliant.

\paragraph{Money transfer in $\BAMP[\emptyset]$}
The main differences between money transfer in $\CAMP[\emptyset]$ and
$\BAMP[\emptyset]$ lies in the fact that a faulty process can try to
transfer money it does not have, and try to present different
behaviors with respect to different correct processes.  This means
that, while the notion of a local history $L_i$ is still meaningful
for a non-Byzantine process, it is not for a Byzantine process. For a
Byzantine process, we therefore define a \emph{mock local history} for
a process $p_i$ as any sequence of transfer operations from $p_i$'s
account\footnote{Let us remind that the operations $\balance()$ issued
  by a Byzantine can return any value. So they are not considered in
  the mock histories associated with Byzantine processes.}.  In this
definition, the mock local history $L_i$ associated with a Byzantine
process $p_i$ is not necessarily the local history it produced, it is
only a history that it could have produced from the point of view of
the correct processes.  The correct processes implement a money-transfer 
object if they all behave in a manner consistent with the
same set of mock local histories for the Byzantine processes.  More
precisely, we define a \emph{mock} history associated with an
execution on a money transfer object in $\BAMP[\emptyset]$ as
$\tilde{H}=(\tilde{L}_1,...,\tilde{L}_n)$ where:
$$
\tilde{L}_j= \begin{cases}
  L_j\text{ if } p_j \text{ is correct, } \\
  \text{a \emph{mock local history} if } p_j \text{ is Byzantine.}\\
\end{cases} 
$$
Considering the  $\BAMP[\emptyset]$ model, 
a money transfer object is such that, for each
of its executions, there exists a \emph{mock} history 
$\tilde{H}$ such that   for any correct process $p_i$, there
is an MT-compliant serialization $S_i$ of  $A_{i,T}(\tilde{H})$.
An algorithm  implementing such executions is said to be MT-compliant.


\paragraph{Concurrent vs sequential specification}
Let us notice that the previous specification considers money transfer
as a concurrent object.  More precisely and differently from previous
specifications of the money transfer object, it does not consider it as a sequential object for which processes must agree on the very  same total
order on the operations they issue~\cite{HW90}. As a simple example, let us
consider two processes $p_i$ and $p_j$ that independently issue the
transfers ${\trf_i(k,v)}$ and ${\trf_j(k,v')}$ respectively.  The
proposed specification allows these transfers (and many others)
to be seen in different order by different processes.
As far as we know, this is the first specification of money transfer as
a non-sequential object.

\section{A Simple Generic Money Transfer Algorithm}
\label{sec:algo}

This section presents a generic  algorithm implementing a money transfer
object. As already said, its generic dimension lies in 
the underlying  reliable broadcast abstraction  used to disseminate
money transfers to the processes, which depends on the failure model. 

\subsection{Reliable broadcast}

Reliable broadcast provides two operations denoted $\rbroadcast()$ and
$\rdeliver()$. Because a process is assumed to
  invoke reliable broadcast
  each time it issues a money transfer, we use a {\it multi-shot}
  reliable broadcast, that relies on \emph{explicit sequence numbers}
  to distinguish between its different instances (more on this
  below). Following the parlance of~\cite{HT94} we use the
following terminology: when a process invokes
$\rbroadcast({\sn,m})$, we say it ``r-broadcasts the message $m$
with sequence number $\sn$'', and when its invocation of
$\rdeliver()$ returns it a pair $(\sn,m)$, we say it
``r-delivers $m$ with sequence number $\sn$''.  While
definitions of reliable broadcast suited to the crash failure model
and the Byzantine failure model will be given in
Section~\ref{sec:algo-proof-crash} and
Section~\ref{sec:algo-proof-byz}, respectively, we state their
  common properties below.
  
\begin{itemize}
\vspace{-0.2cm}
\item Validity.  This property states that there is no message creation.
To this end, it  relates the outputs (r-deliveries) to
  the inputs (r-broadcasts). Excluding  malicious behaviors, a
  message that is r-delivered has been r-broadcast.
\vspace{-0.2cm}
\item Integrity. This property states that there is no message
  duplication.  
\vspace{-0.2cm}
\item Termination-1. This property states that correct processes
  r-deliver what they broadcast.
\vspace{-0.2cm}
\item Termination-2. This property relates the sets of messages
  r-delivered by different processes.
\end{itemize}
\vspace{-0.2cm}
The Termination properties ensure that all the
correct processes r-deliver the same set of messages, and that this
set includes at least all the messages that they r-broadcast.\\

As mentioned above, sequence numbers are used to identify different
instances of the reliable broadcast.  Instead of using an underlying
FIFO-reliable broadcast in which sequence numbers would be hidden, we
expose them in the input/output parameters of the $\rbroadcast()$ and
$\rdeliver()$ operations, and handle their updates explicitly in our
generic algorithm. This reification\footnote{Reification is the
  process by which an implicit, hidden or internal information is
  explicitly exposed to a programmer.} allows us to capture explicitly
the complete control related to message r-deliveries required by our
algorithm.  As we will see, it follows that the instantiations of the
previous Integrity property (crash and Byzantine models) will
explicitly refer to ``upper layer'' sequence numbers.

We insist on the fact that the reliable broadcast abstraction that the proposed
algorithm depends on does not itself provide the FIFO ordering guarantee.
It only uses sequence numbers to identify the different messages sent by
a process. As explained in the next section, the proposed generic
algorithm implements itself the required FIFO ordering property.

\subsection{Generic money transfer algorithm: local data structures}
As said in the previous section, $\init[1..n]$ is an array of
constants, known by all the processes, such that $\init[k]$ is the
initial value of $p_k$'s account, and 
 a transfer of the quantity $v$ from a process $p_i$
to a process $p_k$ is represented by the pair $\langle k,v \rangle$.
Each process $p_i$ manages the following local variables:

\begin{itemize}
\vspace{-0.3cm}
\item $\sn_i$:  integer variable, initialized to $0$,
  used to generate the sequence numbers
  associated with the transfers issued by $p_i$
  (it is important to notice that the point-to-point FIFO order
  realized with the sequence numbers is the 
   only ``causality-related'' control information used in
  the algorithm).
\vspace{-0.25cm}
\item $\del_i[1..n]$: array initialized to $[0,\cdots,0]$
  such that $del_i[j]$ is the sequence  number of the last transfer
  issued by $p_j$ and locally processed by $p_i$.
\vspace{-0.25cm}
\item $\account_i[1..n]$: array, initialized to $\init[1..n]$, that is
  a local approximate representation of the abstract array
  $\Account[1..n]$, i.e., $\account_i[j]$ is the value of $p_j$'s
  account, as known by $p_i$.
\end{itemize}

While other local variables containing bookkeeping information can be
added according to the application's needs, it is important to insist
on the fact that the proposed algorithm needs only the three previous
local variables (i.e., $(2n+1)$ local registers) to solve the
synchronization issues that arise in fault-tolerant money transfer.

\subsection{Generic money transfer algorithm: behavior of a process~$p_i$}
Algorithm~\ref{algo-generic} describes the behavior of a process $p_i$. 
When it invokes $\balance_i(j)$, $p_i$  returns the current value of
$\account_i[j]$ (line~\ref{GMT-01}). 

\begin{algorithm}[h]
\centering{
\fbox{
\begin{minipage}[t]{150mm}
\footnotesize
\renewcommand{\baselinestretch}{2.5}
\resetline
\begin{tabbing}
 aA\=aaA\=aaaaaaaaaaaaA\kill

 {\bf init}: 
 $\account_i[1..n]\leftarrow \init[1..n]$;
 $\sn_i \leftarrow 0$; $\del_i[1..n] \leftarrow [0,\cdots, 0]$.\\~\\

{\bf opera}\={\bf tion} $\balance(j)$ {\bf is}\\
\line{GMT-01} \> $\return(\account[j])$.\\~\\

{\bf operation} $\transfer(j,v)$ {\bf is}\\

\line{GMT-02} \> {\bf if} \= ($v\leq \account_i[i]$)\\

\line{GMT-03} \>\> {\bf then} \= $\sn_i \leftarrow \sn_i +1$;
                $\rbroadcast$$(\sn_i$, {\sc transfer}$\langle j,v\rangle)$;\\

\line{GMT-04} \>\>\> $\wwait~(\del_i[i]=\sn_i)$; $\return(\commit)$\\

\line{GMT-05}  \>\> {\bf else}\> $\return(\abort)$ \\

\line{GMT-06} \> {\bf end if}.\\~\\

{\bf when} $(\sn,${\sc transfer}$\langle k,v\rangle)$ {\bf is}
$\rdelivered$ {\bf from} $p_j$ {\bf  do}\\

\line{GMT-07}\> $\wwait\big((\sn=del_i[j]+1)\wedge(\account_i[j]\geq v)\big)$;\\ 

\line{GMT-08}
\> $\account_i[j]\leftarrow \account_i[j] -v$; 
   $\account_i[k]\leftarrow \account_i[k] +v$; \\

\line{GMT-09}\>  $del_i[j]\leftarrow \sn$.

\end{tabbing}
\normalsize
\end{minipage}
}
\caption{Generic broadcast-based money transfer algorithm
    (code for $p_i$)}
\label{algo-generic} 
}
\end{algorithm}

When it invokes $\transfer(j,v)$, $p_i$ first checks if it has enough
money in its account (line~\ref{GMT-02}) and returns $\abort$ if it
does not (line~\ref{GMT-05}).  If process $p_i$
has enough money, it computes
the next sequence number $\sn_i$ and r-broadcasts the pair
$(\sn_i,$ {\sc transfer}$\langle j,v\rangle)$ (line~\ref{GMT-03}).
Then $p_i$ waits until it has locally processed this transfer
(lines~\ref{GMT-07}-\ref{GMT-09}),  and finally returns $\commit$.
Let us notice
that the predicate at line~\ref{GMT-07} is always satisfied when
$p_i$ r-delivers a transfer message it has r-broadcast.

When $p_i$ r-delivers a pair $(\sn,$ {\sc transfer}$\langle
k,v\rangle)$ from a process $p_j$, it does not process it immediately.
Instead, $p_i$ waits until (i) this is the next message it has to
process from $p_j$ (to implement FIFO ordering) and (ii) its local
view of the money transfers to and from $p_j$ (namely the current
value of $\account_i[j]$) allows this money transfer to occur
(line~\ref{GMT-07}).  When this happens, $p_i$ locally registers the
transfer by moving the quantity $v$ from $\account_i[j]$ to
$\account_i[k]$ (line~\ref{GMT-08}) and increases $\del_i[j]$
(line~\ref{GMT-09}).

\section{Crash Failure Model: Instantiation and Proof}
\label{sec:algo-proof-crash}

This section presents first the crash-tolerant reliable broadcast
abstraction whose operations instantiate the $\rbroadcast()$ and
$\rdeliver()$ operations used in the generic algorithm. Then, using
the MT-compliance notion, it proves that
Algorithm~\ref{algo-generic} combined with a crash-tolerant
  reliable broadcast implements a money
transfer object in $\CAMP[\emptyset]$. It also shows that, in this
model, money transfer is weaker than the construction of an atomic
read/write register.  Finally, it presents a simple weakening of the
FIFO requirement that works in the $\CAMP [\emptyset]$ model.

\subsection{Multi-shot
  reliable broadcast abstraction in  $\CAMP[\emptyset]$}
This communication abstraction, named CR-Broadcast, is defined by the two
operations  $\crbroadcast()$ and $\crdeliver()$.  Hence, we use
the terminology ``to cr-broadcast a message'', and ``to cr-deliver a message''.
\begin{itemize}
  \vspace{-0.2cm}
\item  CRB-Validity.
  If a process $p_i$  cr-delivers a message with sequence number $\sn$
 from a process $p_j$, then $p_j$ cr-broadcast it with sequence number $\sn$. 
  \vspace{-0.2cm}
\item CRB-Integrity. For each sequence number $\sn$ and sender $p_j$ a
  process $p_i$ cr-delivers at most one message with sequence number
  $\sn$ from $p_j$.
  \vspace{-0.2cm}
\item  CRB-Termination-1.
  If a correct  process  cr-broadcasts a message, it cr-delivers it. 
  \vspace{-0.2cm}  
\item CRB-Termination-2.  If a process cr-delivers a message from a
  (correct or faulty) process $p_j$, then all correct processes
  cr-deliver it.
\end{itemize}
CRB-Termination-1 and CRB-Termination-2 capture the ``strong''
reliability property of CR-Broadcast, namely: all the correct
processes cr-deliver the same set $S$ of messages, and this set
includes at least the messages they cr-broadcast. Moreover, a faulty
process cr-delivers a subset of $S$.
Algorithms implementing the CR-Broadcast abstraction in
$\CAMP[\emptyset]$ are described in~\cite{HT94,R18}.

\subsection{Proof of the  algorithm in $\CAMP[\emptyset]$ }
\label{sec:crash-proof}

\begin{lemma}\label{crash-termination}
  Any invocation of $\balance()$ or $\transfer()$ issued by a
  correct process  terminates. 
\end{lemma}
\begin{proofL}
The fact that any invocation of $balance()$ terminates follows
immediately from the code of the operation.

When a process $p_i$ invokes $\transfer(j,v)$, it r-broadcasts a
message and,
 due to the CRB-Termination properties,
$p_i$ receives its own transfer message and the predicate
(line~\ref{GMT-07}) is necessarily satisfied. This is because
 (i) only $p_i$ can transfer its own money,  (ii) the
  $\wwait{}$ statement of line~\ref{GMT-04} ensures the
  current invocation of $\transfer(j,v)$ does not return until the
  corresponding {\sc transfer} message is processed at
  lines~\ref{GMT-08}-\ref{GMT-09}, and  (iii) the fact that
  $\account_i[i]$ cannot decrease
between the execution of line~\ref{GMT-03} and the one of
line~\ref{GMT-07}. It follows that $p_i$ terminates its invocation of
$\transfer(j,v)$.  \renewcommand{\toto}{crash-termination}
\end{proofL}

\noindent
The safety proof is more involved. It consists in showing that any execution 
satisfies MT-compliance as defined in Section~\ref{sec:money-transfer}. 

\vspace{-0.2cm}
\paragraph{Notation and definition}
\begin{itemize}
\vspace{-0.2cm}
\item Let $\trf_j^{\sn}(k,v)$ denote the operation $\trf(k,v)$ issued by $p_j$
 with sequence number $\sn$. 
\vspace{-0.2cm}
\item We say a process $p_i$ {\em processes} the transfer $\trf_j^{\sn}(k,v)$ 
  if, after it  cr-delivered the associated message
  {\sc transfer}$\langle k,v\rangle$ with sequence number $\sn$,
  $p_j$ exits the  wait statement at line~\ref{GMT-07} and executes
  the associated statements at lines~\ref{GMT-08}-\ref{GMT-09}.
  The moment at which these lines are executed is referred to as the
  {\it moment   when the transfer is processed} by $p_i$. (These notions are 
   related to the progress of processes.)
\vspace{-0.2cm}
\item If the message {\sc transfer} cr-broadcast by a process is
  cr-delivered by a correct process, we say that the transfer is
 {\it successful}.  (Let us notice that a message cr-broadcast by a
  correct process is always successful.)
\end{itemize}

\begin{lemma}\label{crash-lemma-1}
If a process $p_i$ processes $\trf_\ell^{\sn}(k,v)$, then any correct
process processes~it. 
\end{lemma} 

\begin{proofL}
Let $m_1,~m_2, ...$ be the sequence of transfers processed by $p_i$
and let $p_j$ be a correct process. We show by induction on $z$ that,
for all $z$, $p_j$ processes all the messages $m_1,~m_2, ...,m_z$.

Base case $z=0$. As the sequence of transfers is empty, the proposition
is trivially satisfied. 

Induction. Taking $z\geq 0$, suppose $p_j$ processed all the transfers
$m_1,~m_2, ...,m_z$. We have to show that $p_j$ processes $m_{z+1}$.
Note that $m_1,~m_2, ...,m_z$ do not typically originate from the same
sender, and are therefore normally processed by $p_j$ in a different
order than $p_i$, possibly mixed with other messages. This also
applies to $m_{z+1}$. If $m_{z+1}$ was processed by $p_j$ before
  $m_z$, we are done.  Otherwise there is a time $\tau$ at which $p_j$
  processed all the transfers $m_1,~m_2, ...,m_z$ (case assumption),
  cr-delivered $m_{z+1}$ (CBR-Termination-2 property), but has not yet
  processed $m_{z+1}$.
  Let $m_{z+1}= \trf_\ell^{\sn}(k,v)$. At time $\tau$, we have
  the following.
\begin{itemize}
\vspace{-0.2cm}
\item On one side, $\del_j[\ell]\leq \sn-1$ since messages
  are processed in FIFO order and $m_{z+1}$ has not yet been
  processed.  On the other side, $\del_j[\ell]\geq \sn-1$
  because either $\sn=1$ or $\trf_\ell^{\sn-1}(-,-) \in m_1,
  ...,m_z$, where $\trf_\ell^{\sn-1}(-,-)$ is the transfer issued by
  $p_\ell$ just before $m_{z+1}= \trf_\ell^{\sn}(k,v)$ (otherwise
  $p_i$ would not have processed $m_{z+1}$ just after
  $m_1,...,m_z$).  Thus $\del_j[\ell]= \sn-1$.
\vspace{-0.2cm}
\item
  Let us now shown that, at time $\tau$, $\account_j[\ell]\geq v$.  To
  this end let $\plus_i^{z+1}(\ell)$ denote the money transferred to
  $p_\ell$ as seen by $p_i$ just before $p_i$ processes $m_{z+1}$, and
  $\minus_i^{z+1}(\ell)$ denote the money transferred from $p_\ell$ as
  seen by $p_i$ just before $p_i$ processes $m_{z+1}$.  Similarly, let
  $\plus_j^{z+1}(\ell)$ denote the money transferred to $p_\ell$ as
  seen by $p_j$ at time $\tau$ and $\minus_j^{z+1}(\ell)$ denote the
  money transferred from $p_\ell$ as seen by $p_j$ at time $\tau$.
  Let us consider the following sums:
\begin{itemize}
\vspace{-0.1cm}
\item
  On the side of the money transferred to $p_\ell$ as seen by $p_j$.
Due to induction, all the transfers to $p_\ell$  included in
$m_1,~m_2,\dots,m_z$ (and possibly more transfers to $p_\ell$)
have been processed by $p_j$, thus
$\plus_j^{z+1}(\ell)\ge \Sigma_{\trf_{k'}(\ell,w)\in \{m_1,m_2, ...,m_z\}} w$
and, as   $p_i$ processed the messages in the order
$m_1, ...,m_z,m_{z+1}$ (assumption), we have 
$\plus_i^{z+1}(\ell)=\Sigma_{\trf_{k'}(\ell,w)\in \{m_1,m_2, ...,m_z\}}w$.
Hence, $\plus_j^{z+1}(\ell)\ge\plus_i^{z+1}(\ell)$. 
\vspace{-0.1cm}
\item
 On the side of the money transferred from  $p_\ell$ as seen by $p_j$.
 Let us observe that $p_j$ has processed all the transfers from
   $p_\ell$ with a sequence number smaller than $\sn$ and no
  transfer from $p_\ell$ with a sequence number greater than or equal to
  $\sn$, thus we have  $\minus_j^{z+1}(\ell)=
  \Sigma_{\trf_\ell(k',w)\in \{m_1,m_2,...,m_z\}} w =\minus_i^{z+1}(\ell)$.
\end{itemize}
Let  $\account_i^{z+1}[\ell]$ be the value of $\account_i[\ell]$
just before $p_i$ processes $m_{z+1}$,
and $\account_j^{z+1}[\ell]$ be the value of $\account_j[\ell]$ at time $\tau$.
As $\account_j^{z+1}[\ell]=\init[\ell] + \plus_j^{z+1}(\ell)-\minus_j^{z+1}(\ell)$
and
$\account_i^{z+1}[\ell]=\init[\ell] + \plus_i^{z+1}(\ell)-\minus_i^{z+1}(\ell)$, 
it follows that $\account_j[\ell]$  is greater than or equal to
the value of  $\account_i[\ell]$ just before $p_i$ processes  $m_{z+1}$,
which was itself greater than or equal to $v$ (otherwise $p_i$ would not have
processed  $m_{z+1}$ at that time). It follows that $\account_j[\ell]\geq v$. 
\end{itemize}
The two predicates of line~\ref{GMT-07} are therefore satisfied, and
will remain so until $m_{z+1}$ is processed (due to the FIFO order on
transfers issued by $p_\ell$), thus ensuring that process $p_j$
processes the transfer $m_{z+1}$.  
\renewcommand{\toto}{crash-lemma-1}
\end{proofL}

\begin{lemma}\label{crash-lemma-2}
  If a  process  $p_i$  issues a successful money transfer  $\trf_i^{\sn}(k,v)$
  (i.e., it cr-broadcasts it in line~{\em\ref{GMT-03}}),
  any correct process  eventually cr-delivers and processes~it.
\end{lemma} 

\begin{proofL}
When process $p_i$ cr-broadcast money transfer
$\trf_i^{\sn}(k,v)$, the local predicate
$(sn=del_i[i]+1)\wedge(\account_i[i]\geq v)$ was true at $p_i$. When
$p_i$ cr-delivers its own transfer message, the predicate is still true
at line~\ref{GMT-07} and $p_i$ processes its transfer  (if $p_i$
crashes after having cr-broadcast the transfer and before processing
it, we extend its execution---without loss of correctness---by assuming
it crashed just after processing the transfer). It follows from
Lemma~\ref{crash-lemma-1} that any correct process processes
$\trf_i^{\sn}(k,v)$.
\renewcommand{\toto}{crash-lemma-2}
\end{proofL}

\vspace{-0.3cm}
\begin{theorem}\label{crash-theorem-1}
Algorithm~{\em \ref{algo-generic}} instantiated with  {\em CR-Broadcast}
implements a money transfer object in the $\CAMP[\emptyset]$ system model,
and ensures that all operations by correct processes terminate.
\end{theorem}
\vspace{-0.1cm}
\begin{proofT}
Lemma~\ref{crash-termination}  proved that  the invocations of the
operations $\balance()$ and $\transfer()$ by the correct processes terminate. 
Let us now consider MT-compliance. 
  
Considering any execution of the algorithm, captured as history
$H=(L_1,...,L_n)$,  let us first consider a
correct process $p_i$. Let $S_i$ be the sequence of the following
events happening at $p_i$ (these events are ``instantaneous''
in the sense $p_i$ is not interrupted when it produces each of them):
\begin{itemize}
\vspace{-0.2cm}
\item the event $\blc(j)/v$ occurs when
  $p_i$ invokes $\balance(j)$ and obtains $v$  (line~\ref{GMT-01}), 
\vspace{-0.8cm}
\item and  the event $\trf_j^{\sn}(k,v)$ occurs when $p_i$ processes the
  corresponding transfer (lines~\ref{GMT-08}-\ref{GMT-09} executed
  without interruption).
\end{itemize}
\vspace{-0.1cm}
We show that $S_i$ is an MT-compliant serialization of $A_{i,T}(H)$.
When considering the construction of $S_i$, we have the following:
\begin{itemize}
\vspace{-0.2cm}
\item
  For all $\trf_j^{\sn}(k,v)\in L_j$ we have that
  $p_j$ cr-broadcast this transfer and that
   $(\sn,$ {\sc transfer}$\langle k,v\rangle)$ was received by $p_j$
  and was therefore \emph{successful}: it
follows from Lemma~\ref{crash-lemma-2} that $p_i$ processes this money
transfer, and consequently we have $\trf_j^{\sn}(k,v) \in S_i$.
\vspace{-0.2cm}
\item
 For all $\op1=\trf_j^{\sn}(k,v)$ and $\op2=\trf_j^{\sn'}(k',v')$
  in $S_i$ (two transfers
issued by $p_j$)  such that $\op1 \rightarrow_j \op2$, we have
$\sn < \sn'$. Consequently $p_i$ processes $\op1$ before $\op2$,
and we have $\op1 \rightarrow_{S_i} \op2$. 
\vspace{-0.2cm}
\item
  For all pairs  $\op1$ and $\op2$ belonging to $L_i$,
  their serialization order is the same in $L_i$ and $S_i$.
\end{itemize}
\vspace{-0.1cm}
It follows that $S_i$ is a serialization of $A_{i,T}(H)$. Let us now
show that $S_i$ is MT-compliant.
\vspace{-0.2cm}
\begin{itemize}
\vspace{-0.2cm}
\item
Case where the event in $S_i$ is $\trf_j^{\sn}(k,v)$. In this case we
have $v\leq \acc(j,\{ \op \in S_i~|~ \op\rightarrow_{S_i} \trf_j(k,v)\}$
because this condition is directly encoded at $p_i$ in the waiting
predicate that precedes the processing of $\op$.
\vspace{-0.2cm}
\item
Case where  the event in $S_i$ is  $\blc(j)/v$. In this case
we have $v=\acc(j, \{ \op \in S_i~|~ \op\rightarrow_{S_i} \blc(j)/v\}$,
because this is exactly the way how the returned value $v$ is computed
in the algorithm. 
\end{itemize}
This terminates the proof for the correct processes.\\

For a process $p_i$ that crashes,
the sequence of money transfers
from  a process $p_j$ that is  processed by $p_i$ is a prefix of
the sequence of money  transfers issued by $p_j$
(this follows from the FIFO processing order, line~\ref{GMT-07}). 
Hence, for each process $p_i$ that crashes 
there is a history $H'=(L_1',...,L_n')$ 
where $L_j'$ is a prefix of $L_j$ for each $j \neq i$ and 
$L_i'=L_i$, such that, following the same reasoning,
the construction $S_i$ given above is an MT-compliant
serialization of $A_{i,T}(H')$,  which concludes the proof of the theorem. 
\renewcommand{\toto}{crash-theorem-1}
\end{proofT}

\subsection{Money transfer vs read/write registers
in  $\CAMP[\emptyset]$} It is shown in~\cite{ABD95} that it is
impossible to implement an atomic read/write register in the
distributed system model~$\CAMP[\emptyset]$, i.e., when, in addition
to asynchrony, any number of processes may crash.  On the positive
side, several algorithms implementing such a register
in~$\CAMP[t<n/2]$ have been proposed, each with its own features (see
for example~\cite{A00,ABD95,MR16} to cite a few).
  An atomic
read/write register can be built from safe or regular
registers\footnote{Safe and regular registers
  were introduced introduced in~\cite{L86}. They 
  have weaker specifications than atomic registers.}~\cite{L86,R12,T06}.
Hence, as atomic registers, safe and
regular registers cannot be built in
$\CAMP[\emptyset]$ (although they can in~$\CAMP[t<n/2]$).
As $\CAMP[t<n/2]$ is a more constrained model
than~$\CAMP[\emptyset]$, it follows that, from a $\CAMP$ computability
point of view, the construction of a safe/regular/atomic read/write
register is a stronger problem than money transfer.

\subsection{Replacing FIFO  by a weaker ordering 
in  $\CAMP[\emptyset]$}
An interesting question is the following one: is FIFO ordering
necessary to implement money transfer in the $\CAMP[\emptyset]$ model?
While we conjecture it is, it  appears that, a small change in the
specification of money transfer allows us to use a weakened 
FIFO order, as shown below.

\paragraph{Weakened money transfer specification}
The change in the specification presented in
Section~\ref{sec:money-transfer} concerns the definition of the
serialisation $S_i$ associated with each process $p_i$. In this
modified version the serialization $S_i$ associated with each process
$p_i$ is no longer required to respect the process order on the
operations issued by $p_j$, $j\neq i$.
This means that two different process $p_i$ and $p_k$ may observe
the $\transfer()$ operations issued by a process $p_j$ in different orders 
(which captures the fact that some transfer operations by a process
$p_j$ are commutative with respect to its current account).

\paragraph{Modification of  the algorithm}
Let $k$ be a constant integer $\geq 1$.
Let $sn_i(j)$ be the highest sequence number such that all the transfer
messages from $p_j$ whose sequence numbers belong to $\{1,\cdots,.sn_i(j)\}$
have been cr-delivered and processed by a certain process $p_i$ (i.e.,
lines~\ref{GMT-08}-\ref{GMT-09} have been executed for these messages).
Initially we have $sn_i(j)=0$.

Let $sn$ be the sequence number of a message cr-delivered by $p_i$ from
$p_j$.  At line~\ref{GMT-07} the predicate $sn=\del_i[j]+1$ can be
replaced by the predicate $sn\in \{sn_i(j)+1,\cdots, sn_i(j)+k\}$. Let us
notice that this predicate boils down to $sn=\del_i[j]+1$ when $k=1$.
More generally the set of sequence numbers $\{ sn_i(j)+1,\cdots,
sn_i(j)+k\}$ defines a sliding window for sequence numbers which allows
the corresponding messages to be processed.

The important point here is the fact that messages can be processed in
an order that does not respect their sending order as long as all the
messages are processed, which is not guaranteed when $k=+\infty$.
Assuming $p_j$ issues an infinite number of transfers, if $k=+\infty$
it is possible that, while all these messages are cr-delivered by
$p_i$, some of them are never processed at
lines~\ref{GMT-08}-\ref{GMT-09} (their processing being always delayed
by other messages that arrive after them).  The finiteness of the
value $k$ prevents this unfair message-processing order from occurring.

The proof of Section~\ref{sec:crash-proof} must be appropriately
adapted to show that this modification implements the weakened money-transfer 
specification.

\section{Byzantine Failure Model: Instantiation and Proof}
\label{sec:algo-proof-byz}

This section  presents first the reliable broadcast abstraction 
whose operations instantiate the $\rbroadcast()$ and $\rdeliver()$
operations used in the generic algorithm. Then, it  proves that
the resulting algorithm correctly implements a money transfer object in
$\BAMP[t<n/3]$.

\subsection{Reliable broadcast abstraction in $\BAMP[t<n/3]$}
The communication abstraction, denoted BR-Broadcast, was introduced
in~\cite{B87}. It is defined by two operations denoted
$\brbroadcast()$ and $\brdeliver()$ (hence we use the terminology
``br-broadcast a message'' and ``br-deliver a message'').  The
difference between this communication abstraction and CR-Broadcast
lies in the nature of failures. Namely, as a Byzantine process can
behave arbitrarily, CRB-Validity, CRB-Integrity, and CRB-Termination-2
cannot be ensured.  As an example, it is not possible to ensure that
if a Byzantine process br-delivers a message, all correct processes
br-deliver it.  BR-Broadcast is consequently defined by the following
properties.   Termination-1 is the same in both
communication abstractions, while Integrity, Validity and Termination-2 consider
only correct processes (the difference lies in the added constraint
written in italics).
\begin{itemize}
  \vspace{-0.2cm}
\item BRB-Validity.  If a {\it correct} process $p_i$ br-delivers a
  message from a {\it correct} process $p_j$ with sequence number
  $\sn$, then $p_j$ br-broadcast it with sequence number $\sn$.
  \vspace{-0.2cm}
\item  BRB-Integrity.
 For each sequence number $sn$ and sender $p_j$ 
 a {\it correct} process $p_i$ br-delivers at most one message with
 sequence number $\sn$ from sender $p_j$.
\vspace{-0.2cm}
\item BRB-Termination-1. If a correct 
  process br-broadcasts a message, it br-delivers it.
\vspace{-0.2cm}  
\item BRB-Termination-2.
If a {\it correct}  process  br-delivers a message from a
(correct or faulty) process $p_j$, then all correct processes br-deliver it. 
\end{itemize}

It is shown in~\cite{BT85,R18} that $t<n/3$ is a necessary requirement
to implement BR-Broadcast. Several algorithms implementing this
abstraction have been proposed. Among them, the one presented
in~\cite{B87} is the most famous. It works in the $\BAMP[t<n/3]$ model,
and requires three consecutive communication steps. The one presented
in~\cite{IR16} works in the more constrained $\BAMP[t<n/5]$ model, but
needs only two consecutive communication steps. These algorithms show
a trade-off between optimal $t$-resilience and time-efficiency.

\subsection{Proof of the algorithm in $\BAMP[t<n/3]$ }
The proof has the same structure, and is nearly the same, as the
one for the process-crash model presented in
Section~\ref{sec:crash-proof}.

\paragraph{Notation and high-level intuition}
 $\trf_j^{\sn}(k,v)$ now denotes a money transfer (or the associated
processing event by a process) that correct processes br-deliver from
$p_j$ with sequence number $\sn$. If $p_j$ is a
correct process, this definition is the same as the one used in the model
$\CAMP[\emptyset]$.  If $p_j$ is Byzantine, {\sc transfer}
  messages from $p_j$ do not necessarily correspond to actual
  $\transfer()$ invocations by $p_j$, but the BRB-Termination-2
  property guarantees that all correct processes br-deliver the
  \emph{same} set of {\sc transfer} messages (with the same sequence
  numbers), and therefore agree on how $p_j$'s behavior should be
  interpreted. The reliable broadcast thus ensures a form of {\it weak
  agreement} among correct processes in spite of Byzantine
  failures. This weak agreement is what allows us to move almost
  seamlessly from a crash-failure model to a Byzantine model, with
  no change to the algorithm, and only a limited adaptation of its proof.

More concretely, Lemma~\ref{crash-lemma-1} (for crash failures)  becomes
the next lemma whose proof is the same  as for Lemma~\ref{crash-lemma-1}
in which the reference to the CBR-Termination-2 property is replaced
by a reference to its  BRB counterpart.
\begin{lemma}\label{byz-lemma-1}
  If a {\em correct} process $p_i$ processes  $\trf_j^{\sn}(k,v)$,
  then any correct process processes it.
\end{lemma}

Similarly, Lemma~\ref{crash-lemma-2} turns into its Byzantine counterpart, lemma~\ref{byz-lemma-2}.

\begin{lemma}\label{byz-lemma-2}
If a {\em correct} process $p_i$ br-broadcasts a money transfer  $\trf_i^{\sn}(k,v)$ (line~{\em\ref{GMT-03}}), any correct processes eventually br-delivers and processes it.

\end{lemma}

\begin{proofL}
When a correct process $p_i$ br-broadcasts a money transfer
$\trf_i^{\sn}(k,v)$, we have $(sn=del_i[i]+1)\wedge(\account_i[i]\geq v)$,
thus when it br-delivers it the predicate of line~\ref{GMT-07} 
is satisfied. By Lemma~\ref{byz-lemma-1}, all the  correct processes 
process this money  transfer. 
\renewcommand{\toto}{byz-lemma-2}
\end{proofL}

\begin{theorem}\label{byz-theorem-1}
Algorithm~{\em \ref{algo-generic}} instantiated with  {\em BR-Broadcast}
implements a money transfer object in the system $\BAMP[t<n/3]$ model,
and ensures that all operations by correct processes terminate.
\end{theorem}

\noindent
The model constraint $t<n/3$ is due only to the fact that
Algorithm~\ref{algo-generic} uses BR-broadcast (for which $t<n/3$ is
both necessary and sufficient).  As the invocations of $\balance()$ by
Byzantine processes may return arbitrary values and do not impact the
correct processes, they are not required to appear in their local
histories. ~\\

\begin{proofT}
 The proof that the operations issued by the correct processes terminate
 is the same as in Lemma~\ref{crash-termination} where the CRB-Termination
 properties are replaced by their BRB-Termination counterparts.

To prove MT-compliance, let us first construct mock local histories for
Byzantine processes: the mock local history $L_i$ associated with a
Byzantine process $p_j$ is the sequence of money transfers from $p_j$
that the correct processes br-deliver from $p_j$ and that they
process.  (By Lemma~\ref{byz-lemma-1} all correct processes process
the same set of money transfers from $p_j$).

Let $p_i$ be a correct process and $S_i$ be the sequence of operations
occurring at $p_i$ defined in the same way as in the crash failure model.
In this construction, the following properties are respected: 
\begin{itemize}
\vspace{-0.2cm}
\item For all, $\trf_j^{\sn}(k,v) \in L_j$ then
\begin{itemize}
\vspace{-0.2cm}
\item
 if $p_j$ is correct, it br-broadcast this money transfer and,
  due to Lemma~\ref{byz-lemma-2}, $p_i$ processes it, hence
  $\trf_j^{\sn}(k,v) \in S_i$.
\vspace{-0.2cm}
\item
if $p_j$ is Byzantine, due to the definition of $L_j$ (sequence of
money transfers that correct processes br-delivers from $p_j$ and
process), we have  $\trf_j^{\sn}(k,v) \in S_i$.
\end{itemize} 
\vspace{-0.4cm}
\item
For all $\op1=\trf_j^{\sn}(k,v)$ and $\op2=\trf_j^{\sn'}(k',v')$ (two
transfers in $L_j\subseteq S_i$) such that $\op1
\rightarrow_j \op2$, we have $\sn < \sn'$,  consequently $p_i$
processes $\op1$ before $\op2$, and we have $\op1 \rightarrow_{S_i} \op2$.
\vspace{-0.2cm}
\item
For all both $\op1$ and $\op2$ belonging to $L_i$,
their serialization order is the same in $L_i$ as in $S_i$ (same as for the
crash case).
\end{itemize}
It follows that $S_i$ is a serialization of $A_{i,T}(\tilde{H})$ where
$\tilde{H}=(L_1, ..,L_n)$, $L_i$ being the sequence of its operations if $p_i$
is correct, and a mock sequence of money transfers, if it is
Byzantine. The same arguments that were used in the crash failure
model can be used here to prove that $S_i$ is MT-compliant.  Since all
correct processes observe the same mock sequence of operations $L_j$
for any given Byzantine process $p_j$, it follows that the algorithm
implements an MT-compliant money transfer object in $\BAMP[t<n/3]$.
\renewcommand{\toto}{byz-theorem-1}
\end{proofT}

\subsection{Extending to incomplete Byzantine networks}
An algorithm is described in~\cite{R20} which simulates a fully
connected (point-to-point) network on top of an asynchronous Byzantine
message-passing system in which, while the underlying communication
network is incomplete (not all the pairs of processes are connected by
a channel), it is $(2t+1)$-connected (i.e., any pair of processes is
connected by $(2t+1)$ disjoint paths\footnote{``Disjoint'' means that,
  given any pair of processes $p$ and $q$, any two paths connecting
  $p$ and $q$ share no process other than $p$ and $q$.  Actually, the
  $(2t+1)$-connectivity is required only for any pair of correct
  processes (which are not known in advance).}).  Moreover, it is
shown that this connectivity requirement is both necessary and
sufficient.\footnote{This algorithm is a simple extension to
  asynchronous systems of a result first established in~\cite{D82} in
  the context of synchronous Byzantine systems.}

Hence, denoting $\BAMP[t<n/3, (2t+1)\mbox{-connected}]$ such a system
model, this algorithm builds $\BAMP[t<n/3]$ on top
$\BAMP[t<n/3,(2t+1)\mbox{-connected}]$ (both models have the same
computability power).  It follows that the previous money-transfer
algorithm works in incomplete $(2t+1)$-connected asynchronous
Byzantine systems where $t<n/3$.

\section{Conclusion}
\label{sec:conclusion}
The article has revisited the synchronization side of the 
money-transfer problem in failure-prone asynchronous
message-passing systems. It has presented a generic algorithm
that solves money transfer in asynchronous message-passing
systems where  processes may experience failures.
This algorithm uses an underlying reliable broadcast communication
abstraction, which differs according to the type of failures
(process crashes or Byzantine behaviors) that processes can experience. 

In addition to its genericity (and modularity), the proposed
algorithm is surprisingly simple\footnote{Let us recall that, in
  sciences, simplicity is a first class property~\cite{AZ10}.
  As stated by A. Perlis --- recipient of the first Turing Award ---
  ``Simplicity does not precede complexity,  but follows it''.}
and particularly efficient (in addition to money-transfer data, each
message generated by the algorithm only  carries  one sequence number).
As a side effect, this algorithm has shown that, in the crash failure
model, money transfer is a weaker problem than the construction of a
read/write register. As far as the Byzantine failure model is
concerned, we conjecture that $t<n/3$ is a necessary requirement for
money transfer (as it is for the construction of a read/write
register~\cite{IRRS16}).

Finally, it is worth noticing that this article adds one more member
to the family of algorithms that strive to ``unify'' the crash failure
model and the Byzantine failure model as studied
in~\cite{BN91,DG16,IRS16,NT90}.

\section*{Acknowledgments}
This work was partially supported by the French ANR projects
16-CE40-0023-03 DESCARTES, devoted to layered and modular structures in
distributed computing, and ANR-16-CE25-0005-03 O'Browser, devoted to decentralized applications on browsers.


\end{document}